\documentclass[11pt]{article}
\usepackage{graphicx}
\usepackage{wrapfig}

\usepackage{amssymb}

\begin{document}

\baselineskip=24pt

\begin{titlepage}

\centerline{\bf Validation in Fusion Research:}
\centerline {\bf Towards Guidelines and Best Practices}

\bigskip

\centerline{P.W. Terry$^{a)}$, M. Greenwald$^{b)}$, J.-N. Leboeuf$^{c)}$, G.R. McKee$^{a)}$,}
\centerline{D.R. Mikkelsen$^{d)}$, W.M. Nevins$^{e)}$, D.E. Newman$^{f)}$, and D.P. Stotler$^{d)}$}

\medskip

\centerline{\it Task Group on Verification and Validation}
\centerline{\it US Burning Plasma Organization and US Transport Task Force}

\bigskip

\noindent{$^{a)}$}University of Wisconsin-Madison, Madison, Wisconsin

\noindent{$^{b)}$}Plasma Science and Fusion Center, MIT, Cambridge, Massachusetts

\noindent{$^{c)}$}University of California at Los Angeles, Los Angeles, California

\noindent{$^{d)}$}Princeton Plasma Physics Laboratory, Princeton, New Jersey

\noindent{$^{e)}$}Lawrence Livermore National Laboratory, Livermore, California

\noindent{$^{f)}$}University of Alaska at Fairbanks

\begin{abstract}

\baselineskip=24pt

Because experiment/model comparisons in magnetic confinement fusion have not yet satisfied the requirements for validation as understood broadly, a set of approaches to validating mathematical models and numerical algorithms are recommended as good practices.  Previously identified procedures, such as verification, qualification, and analysis of error and uncertainty, remain important.  However, particular challenges intrinsic to fusion plasmas and physical measurement therein lead to identification of new or less familiar concepts that are also critical in validation.  These include the primacy hierarchy, which tracks the integration of measurable quantities, and sensitivity analysis, which assesses how model output is apportioned to different sources of variation. The use of validation metrics for individual measurements is extended to multiple measurements, with provisions for the primacy hierarchy and sensitivity.   This composite validation metric is essential for quantitatively evaluating comparisons with experiments.  To mount successful and credible validation in magnetic fusion, a new culture of validation is envisaged.

\end{abstract}

\end{titlepage}

\setcounter{page}{2}

\noindent{\bf I. INTRODUCTION}

\medskip

Predictive capability has emerged as a key goal in magnetic confinement fusion research, not just for its special value in designing and operating ever costlier and more complex devices such as ITER \cite{ITER} and DEMO \cite{DEMO}, but, more generally, because its attainment would signify the quantitative maturity in understanding and modeling that is required for success in fusion.  
Genuine predictive capability  will require computational models that have been shown to be valid under widely accepted standards.  This paper identifies and explores issues that must be confronted in demonstrating the validity of computational models in fusion.  
It provides a starting point on the path toward a community consensus on what validity means by proposing guidelines and good practices in validation of computational models.

Verification and Validation (V\&V) have been formulated in related scientific communities, such as fluid dynamics, to codify a certification procedure necessary for predictive capability \cite{ober02}.  Verification is the process by which it is determined that a numerical algorithm correctly solves a mathematical model within a set of specified, predetermined tolerances.  Validation is the process by which it is determined that the mathematical model faithfully represents stipulated physical processes, again within prescribed limits.  To a limited extent there is an emerging culture of verification in the US fusion program \cite{ECCTTF07}-\cite{myra07}.  
These and other good faith verification efforts need to be pursued more widely, 
and extended to methodologies formulated for verification by experts in computer sciences and elsewhere, with careful and accessible documentation becoming the norm.
However, as far as we are aware, nothing yet undertaken in the fusion community satisfies the requirements of validation as it is understood in other communities.  What is lacking generally are quantitative assessments of discrepancies between model and experiment and their mapping into a set of tolerances applied to prediction, and the community consensus on standards and procedures that would lead to acceptance of a process as constituting validation.  
Because prescriptions and templates for verification have been formulated in considerable detail, whereas validation is less well prescribed, this paper will focus on validation.

It may not be tenable for the fusion community to pursue V\&V exactly as envisioned in other communities.  While efforts in allied fields serve as critical guideposts, certain realities in fusion research have to be confronted and accommodated.  These include budget and manpower constraints that do not allow the same magnitude of effort.  They include special complexities of modeling that extend beyond the usual problems in turbulence of nonlinearity, multiple scales, and geometry.  For example, the dynamics of fusion plasmas is not described by a single model, but requires complex integration of a series of distinct, physically realizable models with vastly different scales, representations of plasma physics, and computational requirements.  Another challenge includes the routine occurrence of multiple equilibrium states, bifurcation dynamics, and extreme sensitivity from features like strong transport near critical gradients. The fusion plasma environment also poses severe problems for measurement in terms of limited diagnostic capability (many crucial quantities cannot be measured), limited diagnostic access, and the necessity of applying significant a priori modeling to interpret the measurements.  Finally, predictive fusion codes will have their own set of uses to which fusion V\&V will have to be adapted.

\medskip

\noindent{\bf II. FUSION MODELING ISSUES}

\medskip

It is not clear how successful fusion modeling will be in overcoming all of the challenges listed in the preceding paragraph.  On the time scales required to design and produce new machines the success may be partial.  It is also anticipated that on ITER, reduced simulation codes will be used for shot planning and analysis, requiring a tradeoff between the demands for physics fidelity and fast turn around time. 
Therefore codes are validated as representing physical behavior, but within tolerances usually quantified in something called a validation metric.  The tolerances are not to be set a priori as a design parameter, but a posteriori, as a quality grade through the process of comparison with experiment.  The challenges listed above impact the fidelity of models.  Where there is significant uncertainty in experimental measurements, tolerances will be larger.  Likewise where there are model limitations from the standpoint of physics included, resolution achieved, or other factors, larger tolerances will reflect the limitation.   

Present-day efforts at comparison of models and experiment reveal a number of problems intrinsic to the process.  In some sense, validation can be couched as the procedure by which these problems are met, and their consequences are quantitatively assessed and mapped onto tolerances.  
We briefly present some of these problems, and then recommend approaches for meeting them.
 
\noindent{$\bullet$ \it  Discrepancies between model and experiment} - Disagreement between the results of modeling and experimental measurement must be characterized, understood and ultimately quantified.  What are the sources of disagreement?  Which matter more or less?  Is the understanding of discrepancies consistent with trends where agreement becomes better or worse?  How good is ``good'' when it comes to agreement?
The widely followed practice of comparing a model result with experiment and declaring the agreement ``reasonable'' or not, does not constitute validation because it is a qualitative assessment.

\noindent{$\bullet$ \it Fortuitous agreement} - Given the complexity of plasma dynamics and models, the possibility that good agreement may be fortuitous must be confronted.  In cases where agreement has been reached by successively improving the realism of the model, will the addition of other effects thought to be less important destroy the agreement?  Are there other quantities to compare which may show the agreement to be fortuitous or not particularly meaningful?

\noindent{$\bullet$ \it Sensitivity} - Plasmas are highly nonlinear.  The result is that certain quantities are very sensitive to changes in the parameters on which they depend.  Agreement in quantities that are very sensitive may be difficult to achieve, because small discrepancies in underlying parameters translate into large uncertainties in the sensitive quantity.  Quantities that are not sensitive may not be able to discriminate between physically different models.

\noindent{$\bullet$ \it Differentiating between models} - Certain comparisons may have limited value in validation because physically different models (producing different outcomes for other comparisons) tend to produce equally good agreement within some set of tolerances.  It is important to determine what quantities lead to these kinds of comparisons, so that validation is not based solely on them.

\noindent{$\bullet$ \it Optimizing comparisons} - Certain quantities are sensitive, while others lead to comparisons that do not discriminate well.  But quantities for comparison are also affected by diagnostic and computational limitations.  These factors must be assessed in some fashion to optimize comparisons.

\medskip

\noindent{\bf III. VALIDATION APPROACHES}

\medskip

The difficulties discussed above can be attacked through a series of approaches and validation activities.  As indicated in Table 1, these approaches are cross cutting. To differing extents all deal with the problems encountered in making meaningful comparisons with experiment.  In discussing these approaches we assume that they are applied to codes that have undergone a thorough, documented verification process. 

Many of these approaches and ancillary activities have been defined and incorporated into glossaries.
Glossaries have been drafted by the Society for Computer Simulation, the Institute of Electrical and Electronics Engineers, the Defense Modeling and Simulation Organization, and the American Institute of Aeronautics and Astronautics.  These are discussed in Sec. 2.1 of Ref.~\cite{ober02}.  A glossary adapted for fusion simulation, and somewhat oriented toward verification, has been formulated by the European Fusion Development Agreement (EFDA) Task Force on Integrated Tokamak Modelling \cite{strand}.  In considering validation for fusion modeling, we have found it desirable to introduce some new concepts.  Our glossary is given in Appendix A.   We introduce here those concepts most central to a discussion of approaches to validation in fusion systems.

\medskip

\noindent{\bf A. Key Concepts}

\medskip

\smallskip

\noindent{\it Validation} - The process of determining the degree to which a model is an accurate representation of the real world from the perspective of the intended uses of the model; an exercise in physics.

\smallskip

\noindent{\it Qualification} - A theoretical specification of the expected domain of applicability of a conceptual model and/or of approximations made in its derivation.

\smallskip

\noindent{\it Uncertainty} - A potential deficiency in any phase or activity of the modeling process that is due to lack of knowledge.  Uncertainties can arise in the models themselves and/or in the experimental data used for validation.

\smallskip

\noindent{\it Error} - A recognizable deficiency in any phase of activity of modeling and simulation that is not due to lack of knowledge.  Errors may be introduced by insufficient resolution in a simulation, programming bugs in the code, or physics excluded from the model.

\smallskip

\noindent{\it Sensitivity Analysis} - The study of how the variation in the output of a model (numerical or otherwise) can be apportioned, qualitatively or quantitatively, to different sources of variation.

\smallskip

\noindent{\it Primacy Hierarchy} - Ranking of a measurable quantity in terms of the extent to which other effects integrate to set the value of the quantity.  Assesses ability of measurement to discriminate between different non-validated models.

\smallskip

\noindent{\it Validation Metric} - A formula for objectively quantifying a comparison between a simulation result and experimental data.  The metric may take into account errors and uncertainties in both sets of data as well as other factors such as the primacy of the quantities being compared.  

\medskip

We now describe how these concepts are utilized in approaches to validation.

\medskip

\noindent{\bf B. Qualification}

\medskip

Qualification is essential in validation.  It should not be overlooked even though a model might be well established.  For example, the success of MHD in equilibrium modeling supports a widely held perception that the model is predictive.  However, this only applies to equilibrium modeling.  If there are bifurcations, and MHD is used to model the transients, the model must undergo qualification for that application, a process that is far from trivial.  
Similarly, examples of model validation for energetic particle physics can be found for cases with trace levels of alpha particles or lower energy beam particles.  To validate such models at higher alpha particle fractions or beam power levels requires a qualification process that takes note of such limitations, determines whether they are intrinsic to the regime of validity of the model, and traces how model fidelity responds quantitatively to the desired parameter changes.  This includes identifying the extent to which behavior and regimes are linear or nonlinear, and how the validated regimes relate to the regimes that may arise in future experiments.
Therefore, while qualification is characterized as a theoretical exercise, it must be carried out in the context of the experimental plasma conditions under which the model will be applied. Qualification must lead to a quantitative rating of the applicability of a model given its theoretical constraints and the parameters of the experiment to which it will be applied.  This may be straightforward if all experimental parameters lie comfortably within the domain of applicability.  However, if parameters lie close to the edge of applicability, nonlinearity in combining the quantitative contributions of neglected effects makes this a nontrivial exercise.  Assessing these matters quantitatively is part of setting up a validation metric, as described below.

\medskip

\noindent{\bf C. Uncertainty and Deficiencies}

\medskip

An assessment of uncertainty in experiment is often denoted with error bars.  
Error bars frequently reflect only statistical uncertainty.  Nonetheless it is important to document and describe how error bars are obtained.  Ratings of statistical error rely on statistical assumptions whose validity in turn must be established.  For example, sampling restrictions associated with diagnostic limitations may favor Gaussian statistics.  On the other hand, uncertainties due to dynamical chaos can be non Gaussian.  Such issues may affect comparisons.  Systematic error is incorporated less often into error bars, yet it can lead to significant discrepancies with models.  Sources of systematic error include modeling errors and uncertainties in equilibrium solvers, lack of precision in equilibrium-solver inputs, limitations in diagnostic sensitivity and resolution, deconvolution of line-integrated measurements, modeling inherent in diagnostic signal interpretation ({\it e.g.}, inferring density from ion saturation current in Langmuir probes), and additional processing of diagnostic signals in analysis.  Systematic error is obviously more difficult to rate numerically than statistical uncertainty, however it must be undertaken and folded into a validation metric.

Uncertainty and deficiencies in the modeling process arise from a variety of sources.  These include imperfect mappings of magnetic topology to laboratory coordinates and restrictions on equilibrium specification and its modeling.  There are generally limitations arising from physical processes excluded, such as kinetic effects, fluctuation fields, inhomogeneities, and boundary physics.  Uncertainties arise from artificial constraints, including fixed profiles, flux tubes, and missing or imprecise experimental data for input parameters.  Some uncertainty is intrinsic to the algorithm that implements the conceptual model.  These include limitations on resolution, both temporal and spatial, integration time, and artificial numerical dissipation.  Once uncertainties in modeling are assessed, they should be  
tabulated and displayed as expected deviations from computed values at specified confidence levels.  Section III.F gives examples.

It may be possible to circumvent certain limitations and uncertainties in measurement by use of synthetic diagnostics.  Direct comparison of the experimental measurements with processed simulation output may be a useful approach particularly when inversion of experimental measurements to produce local quantities of interest is not possible or accurate.  This situation can arise when there is limited spatial coverage or averaging or because the inversion process is mathematically ill-conditioned.
Synthetic diagnostics incorporate the experimental spatial and temporal transfer functions, mimic diagnostic uncertainties arising from sensitivity and resolution limitations, and replicate plasma modeling inherent in diagnostic signal interpretation \cite{bravenec}.  
However, synthetic diagnostics are themselves models, and need validation.  Validation of synthetic diagnostics is not a trivial issue and must not be overlooked, whether aspects are done independently (e.g., for calibration), or whether validation is done integrally for both the measuring diagnostics and the model whose results are being measured.
Moreover, because synthetic diagnostics integrate and process their input data, they may increase the tendency for fortuitous agreement or make it difficult to differentiate between models.  However, they can be valuable in quantifying experimental uncertainties for inclusion in validation metrics.

\medskip

\noindent{\bf D. Primacy Hierarchy}

\medskip

Not all experimentally measured quantities provide equally meaningful comparisons with models.  
This is inherent in the issues presented in Sec.~II and the errors and uncertainties described above. 
One way of rating measured quantities for comparisons is the primacy hierarchy.  The primacy hierarchy tracks how measured quantities integrate or combine to produce other measured quantities.  A quantity is assigned to a lower primacy level if its measurement integrates fewer quantities.  Note that, as described, the primacy hierarchy is a property of the way a quantity is measured, not the quantity itself.  
We give an example of a primacy hierarchy for transport measurements in which the primary quantities contributing to the particle flux are Fourier amplitudes of density and potential fluctuations, and wavenumber.  The wavenumber and potential fluctuation combine to produce a secondary quantity, the E$\times$B flow.  The flow and density combine to produce a tertiary quantity, the particle flux.  A fourth level is the diffusivity, which combines the flux and a lower level quantity, the profile.  The diffusivity is frequently the basis for comparisons.  It not only integrates three lower levels and any uncertainties or deficiencies present in the measurements, but it requires the application of an approximate transport model.  Consequently, comparisons of diffusivities are not especially useful for validation.  An example of the particle transport hierarchy is given in Fig.~1.  A second example, given in Fig.~2, is drawn from wavenumber spectrum measurements.  It assigns primacy levels to fluctuations, spectrum, and correlation length.

The closeness of agreement between an experiment and a model is generally a function of primacy level.  This is illustrated in Fig.~3, which shows comparisons of a gyrokinetic model and experiment for level-1 density fluctuations (Fig. 3a) and the level-3 flux (Fig. 3b) \cite{ross}.  The code and experiment appear to be in better agreement for the higher level quantity, presumably because deficiencies and uncertainties partially cancel in the integration from level 1 to level 3.  This particular trend with primacy level is common but not universal.  In certain measurements the integration could amplify the disagreement as primacy level increases.  Given such variations, it is not desirable to base validation comparisons on a single primacy level.  Indeed, the issues of integration and sensitivity (which is described in the next section), make it important to use multiple measures {\it both within} a primacy level {\it and across} primacy levels.

For example, a comparison at a single primacy level that tends to optimize agreement between models and experiment may not be able to differentiate between models with substantially different physics (and substantially larger discrepancies at other levels).  This is evident in comparisons of wavenumber spectra made over the years.  Fairly simple drift wave theory produced good agreement with ATC \cite{ATC} fluctuation data in 1976 \cite{horton76}.  A decade later a more sophisticated analytic theory achieved similarly good agreement with PRETEXT \cite{TEXT} data \cite{ter85}.  In 2006 a gyrokinetic code, far more advanced than either analytic theory, again achieved good agreement with Alcator C-Mod \cite{CMOD} data \cite{ern06}.  The sense that wavenumber spectra do not provide stringent model validation is bolstered by other comparisons made in the PRETEXT case.  These showed that the collisional drift wave model of Ref.~\cite{ter85} could not satisfactorily model other aspects of the fluctuation physics.  Thus, meaningful validation will require quantitative knowledge of trends across the primacy hierarchy, as garnered from comparisons at multiple levels.  As part of a validation metric ratings should be assigned to comparisons at different levels in the hierarchy, as described below in Sec. III.F.3.  

\medskip

\noindent{\bf E. Sensitivity Analysis}

\medskip

Magnetically confined plasmas are nonlinear and complex, producing sensitivities in parameters, scalings, and dynamics.  These may be either known or unknown.  Known examples are bifurcations between disparate equilibria controlled by small variations of critical parameters, and stiffness in profiles whereby small changes in gradients lead to large changes in fluctuation levels and transport fluxes.  Not all quantities are equally sensitive.  Sensitivities pose challenges for validation that must be understood and confronted.  Sensitivities amplify uncertainties and deficiencies, potentially leading to large discrepancies in comparisons between model and experiment.  A model validated with tight agreement in non sensitive comparisons could have large discrepancies in a sensitive comparison.     
To understand the uncertainties in a given computational result one must understand how sensitive those results are to the various input parameters.  This is particularly important to investigate over the range of intrinsic uncertainty in those input parameters.

Comparative studies of turbulent transport find a significant sensitivity of fluctuation levels and transport fluxes to profiles.  This sensitivity has a robust theoretical basis in the strong increase of transport as a function of gradient scale length above a critical value.  It is evident in Fig.~3b, which shows fluxes rising by a factor of 4-5 for a 60\% increase in the gradient scale length parameter \cite{ross}.  This sensitivity was cited as the primary reason for the large discrepancy between experiment and model in the comparisons of Fig.~3b.  

It is possible to construct or find measurables or functions of measurables in which the sensitivities apparently cancel out or are not present.  Several such quantities have been proposed.  An example of the former is the ratio of two diffusivities such as an ion heat diffusivity to an electron heat diffusivity.  Examples of the latter include quantities like the wavenumber of the spectrum peak and the radial correlation length.  Finding such quantities is not likely, in itself, to provide a solution to the sensitivity problem, although it may enable falsification of the hypothesis that a model is valid even when strong sensitivities are involved.  
For example the ratio of electron and ion heat heat fluxes is often insensitive to variations in the temperature gradient scale length $L_{T_i}$.  If a model robustly fails to predict $Q_e/Q_i$ it can be assumed that the model is not valid for individual heat fluxes, even when there is insufficient precision to determine whether individual heat fluxes are incorrect.  However, the converse does not hold, {\it i.e.}, success in predicting $Q_e/Q_i$ does not provide validation for the modeling of individual heat fluxes.
This difficulty impacts the capacity to differentiate between different models.  Measurables that have some dependence on a sensitive parameter, but reduce the sensitivity, are likely to also reduce the capacity to differentiate between different models.  Radial correlations apparently have this property, as illustrated by comparisons with experiment and two significantly different models, both of which do quite well in the comparison \cite{rhodesIAEA}.  Sensitivity and primacy hierarchy can be coupled.  Measurables that reduce sensitivity may be at a higher primacy level.  This is true of the ratio of heat diffusivities.  It combines numerator and denominator, raising the primacy level and, in a fairly obvious way, allowing cancelation of sensitivities, deficiencies, and uncertainties to produce closer agreement.  It also introduces the approximate model assumption of diffusive transport.

The above difficulties make it necessary to thoroughly understand and quantify the sensitivities of the model.  This is accomplished with sensitivity analysis, which is performed numerically but is informed in key ways with theory input. 
Theory creates the conceptual framework for describing the physics being modeled, identifying features of the dynamical landscape and the workings of the underlying processes.  Theory provides qualitative and quantitative descriptions in terms of basic scalings, identification of crucial parameters, characterization of sensitivities, and the morphology of dynamical behavior.  An example is the effect of E$\times$B shear, which was posited theoretically, verified experimentally, and then shown to be important for inclusion in models for reducing discrepancies with experiment (but not removing them altogether).  Using theory for guidance, numerical approaches map out, characterize, and quantify all of the sensitivities of the model.  This analysis determines if there are low-sensitivity measurables that are capable of differentiating between different models.  It tests whether there are functions of measurables that reduce sensitivity without reducing deficiencies and uncertainties.  Thorough sensitivity analysis may conclude that reduction of discrepancies in compared measurables may require reduction of uncertainties in source parameters.

\medskip

\noindent{\bf F. Validation Metric}

\medskip

Once a model is encoded, it is a numerical exercise to compute measurable quantities and graphically compare them to experimental observations. A straightforward assessment of the comparison inevitably results in qualitative judgments such as ``pretty good agreement'' or ``significant differences'', {\it etc.} This kind of evaluation has typified comparisons of fusion models and experiments to date.  While this qualitative approach is suitable for testing models during the early stages of model development or demonstrating general trends, validating a model for predictive applications calls for a more quantitative, objective approach. The evaluation must yield a numerical rating of the extent to which the model is validated within a context that assigns a widely understood meaning to that rating.  This is accomplished with a validation metric. 
For example, in the context of prediction, a validation metric might yield a declaration of the maximum differences expected or possible between a prediction and reality in a given measurable quantity for some specified confidence level. To make this statement, one must anticipate the variations expected in the simulation due to changes in the model input parameters that are possible within the uncertainties and deficiencies, both of model and experiment. To provide meaning to the variations characteristic of a given measurable, it is necessary to know how measures perform across the primacy hierarchy and sensitivity landscape. An example of a comparison interpreted in this fashion is given in Fig.~4. 

The validation metric could incorporate quantitative information about 1) approximations and validity regimes identified in the qualification process; 2) uncertainties and deficiencies in experiment, including statistical and systematic error; 3) uncertainties and deficiencies in modeling; 4) location of measured quantities within the primacy hierarchy; and 5) model sensitivity. Generally speaking, 1) - 3) represent the tolerances intrinsic to a comparison, whereas 4) - 5) track the significance and meaning of comparison results. There is no standardized or algorithmic procedure for quantifying these individual assessments. Nor is there any general prescription for combining them into the validation metric. These operations will 
have to be defined through a collective development process subject to the usual trial and error of scientific endeavor. 

\medskip

\noindent{\bf Simple Metrics}

\medskip

General recommendations and approaches can be made, including those that  leverage off practices established in other fields.  For example, the relatively simple validation metric described by \cite{OandB} is a natural step beyond the familiar, qualitative comparison. The key requirement for evaluating this metric is the availability of multiple experiments for each condition being examined (labeled here by $x$).  For illustrative purposes, suppose that experiments have been performed over a range of $x$, with two or more experiments done at a given $x$.  Simulations based on those experiments have also been run over this same range of $x$. The availability of multiple experiments at each $x$ allows mean values and confidence intervals of the observable quantity $\Sigma(x)$ to be computed directly from the data. This approach is preferable to the conventional ``errorbar'' approach since the latter is inevitably based on arguments and models that themselves must be tested before they can be considered reliable. 

The actual validation metric in this example is the absolute value of the difference between the experimental and simulation data, averaged over the distribution of the latter (see Appendix).  The effectiveness of the model, and of the validation exercise itself, is then assessed by comparing this difference with the experimental confidence interval, as in Fig. 5.  For example, the simulation error exceeds the 90\%
confidence interval at the largest $x$ values, where the confidence intervals are relatively small, suggesting that future model improvements would be most 
profitably focused on this regime.  In contrast, the confidence intervals are relatively large for small to medium values of $x$, preventing the trend predicted by the simulation in Fig. 4 from being tested. The conclusion of the exercise would then be that more experiments and/or more precise measurements are needed in this regime. 
Once a model has demonstrated its effectiveness over a range of $x$, the next
level of comparison is to compute a global metric by averaging the relative deviations and confidence intervals over $x$.  This process and the result are described in 
Appendix B.  

\medskip

\noindent{\bf Comparisons Involving Simple Metrics}

\medskip

Validating models is often aided by comparing codes with different approximation schemes.  Methods for the transparent display of such comparisons have been devised, particularly where there is a large number of models.  An example is the Taylor diagram \cite{tay01} which is used in climate modeling to assess the relative performance of general circulation models across many climatological measures.  In a fusion context the Taylor diagram can be used to compare modeled profiles with experimental observation across a variety of models, parameterization schemes, and diagnostic techniques; and for different profiles.  In the Taylor diagram, functions describing the variation of a quantity across some domain are represented by a point on a 2D polar plot \cite{tay01}.  One point is designated as a reference.  It may be one model in a comparison between many models.  If models are being compared to experimental observation for validation, the experimental results provide a sensible reference.  The radial position of each point on the Taylor diagram gives the standard deviation of the function about its mean value.  The angular position, as measured from the horizontal axis, is $\cos^{-1}R$, where $R$ is the correlation coefficient between the reference function and the model function.  The RMS difference with the reference is the distance between a model point and the reference point.  Because it is a simple function of the other two quantities, it is not an independent quantity.  The reduction of profiles to a point allows a single diagram to display the comparison of many different models, variations of inputs associated with uncertainties, or even different measures (reflecting differences across the primacy hierarchy).  

A simple example of a Taylor diagram for fusion comparisons is given in Fig.~6.  This figure shows a tabulation of the variation of correlation length with minor radius for a set of discharges, labeled ``Ref'', and four models - a neoclassical ion temperature gradient (ITG) model, two slab ITG models, and a toroidal ITG model.  The information has been extracted from Fig.~3 of Ref.~\cite{rhodes}.  Because the Taylor diagram of Fig.~6 is a variation of the diagram as described in Ref.~\cite{tay01} we provide the details of the differences.  In Fig.~6 standard deviations are normalized to their respective means in the radial axis of the plot.  If $r_n$ are the discrete values of the reference, {\it e.g.}, values of a measure at different radial positions, and $f_n$ are the corresponding values generated by a computed model, the correlation coefficient $R$ is defined as
\begin{equation}
R=\frac{1}{N}\sum_{n=1}^N(f_n-\bar{f})(r_n-\bar{r})/\sigma_f\sigma_r.
\end{equation}
Here $\bar{f}$ and $\bar{r}$ are the mean values and $\sigma_f$ and $\sigma_r$ are the standard deviations of $f$ and $r$.  The radial distance from each point to the origin is $\hat{\sigma}_f=\sigma_f/\bar{f}$ ($\hat\sigma_r=\sigma_r/\bar{r}$ in the case of the reference).  The angle, measured from the horizontal axis is $\cos^{-1}R$.  The distance between the reference point and each model point is the RMS deviation,
\begin{equation}
\hat{E}'=\Bigg\{\frac{1}{N}\sum_{n=1}^N\Big[\frac{(f_n-\bar{f})}{\bar{f}}-\frac{(r_n-\bar{r})}{\bar{r}}\Big]^2\Bigg\}^{1/2}.
\end{equation}
A scale for the RMS deviation is often placed on the Taylor diagram as concentric arcs centered on the reference point.  These are the broken arcs in Fig.~6.  The relationship among the correlation coefficient, the RMS deviation between reference and model, and the standard deviation is simply the law of cosines,
$\hat{E}'^2=\hat\sigma_f^2+\hat\sigma_r^2-2\hat\sigma_f\hat\sigma_rR$.

This figure reduces the fidelity of models to single points and allows a rapid assessment to guide refinements in modeling.  At a glance it is evident from Fig.~6 that the toroidal ITG comes closest to the normalized standard deviation and has the lowest RMS difference.  One of the slab ITG models has the poorest performance in these areas.  All of the models have a high value of the correlation coefficient.  This indicates that correctly capturing the decrease of radial correlation with minor radius is not particularly challenging for models, and therefore is not a particularly stringent quantitative test.

\medskip

\noindent{\bf Composite Metrics}

\medskip

As discussed earlier, construction of multiple validation metrics at various levels on the primacy hierarchy is very important to give confidence in the overall validation.  
Therefore, after constructing individual validation metrics, a quantitative assessment of the effect of primacy hierarchy and sensitivity, and their factorization into a composite validation metric, is the next important step. 
There is an additional rationale for introducing the primacy hierarchy and sensitivity into a metric.  While it is straightforward to combine multiple but independent measures of the same quantity, it is less obvious how to combine measures of different quantities.  Primacy level and sensitivity, as weights in a combination of measures, provide a method.

The primacy hierarchy and sensitivity have not been widely addressed; hence there is little guidance available for using them in metrics. Ratings for these aspects of a comparison can be incorporated into a confidence level or skill score [15]. As an example of this type of rating we consider a composite validation metric. The construct we offer is admittedly non-unique and other composite metrics can be constructed that are equally valid and might be better suited to some applications. Our composite metric is therefore meant to illustrate one possible form, incorporating the minimum elements as a starting point. The idea is to build an objective, reproducible validation metric that is a composite of individual metrics used to validate a model.  At the same time, the composite metric should be constructed to minimize the possibility of manipulating the metric to some advantage, either by chance or design.  The composite metric will be constructed from an additive combination of individual metrics weighted by their position in the primacy hierarchy and their sensitivity to parameters. This will allow an overall assessment of goodness of validation consistent with the notion that multiple measures across the primacy hierarchy yield a higher grade of validation.

The composite metric is the sum of weighted metrics from individual measures.  The sum allows higher scores for validation exercises combining larger numbers of comparisons with experiment (or other models).   The sum  is expressed as 
\begin{equation}
M_s=\sum_i B_iP_iS_iW_i\frac{1}{10}, 
\end{equation}
where $B_i$ is a normalized rating of the goodness of an individual measure, $P_i$ is a normalized value on the primacy hierarchy, $S_i$ is a normalized sensitivity, and $W_i$ is a repetition weight.  The rating $B_i$ is normalized to a scale of 0 to 1, with values near 0 representing a comparison with large discrepancies, errors and uncertainties, {\it etc.}, and 1 representing a faultless comparison.  Alternatively, this quantity could represent a binary scoring scheme with 0 as fail and 1 as pass.  It could also be a normalized Bayes factor.  The primacy rating $P_i$ is normalized to a scale of 1 to 5, with 5 corresponding to a measure at the lowest level of a primacy hierarchy (least integration) and 1 at the highest level.  The sensitivity weight $S_i$ ranges between 1 and 2, with 1 corresponding to a measure with the lowest sensitivity, and 2 a measure with the highest sensitivity.  The repetition weight $W_i$ prevents multiple measures of a repetitive nature on the same primacy level from artificially raising the score of a summed quantity by virtue of sheer repetition.  For repeated measures, for example, a fluctuation intensity at different spatial locations, the first measure is weighted $W_1 = 1$, the second $W_2= 0.5$, the third $W_3= 0.25$, and the $n$th $W_n=(0.5)^{n-1}$.  The summation, in connection with the repetition weight, obviously rewards diversified measures.  The unbounded composite score $M_s$ rises with the number of such measures.  The higher the score, the higher the skill level of the model.  Experience will show how large $M_s$ might become, but we anticipate on the basis of the types of comparisons that have been done, that a score of 1 or lower would be considered poor, scores in the range of 1 to 5 would be considered reasonable, and scores above 10 would be considered good.  

Because it is possible that inclusion of a sufficiently large number of poorly agreeing measures could result in a good score, a second metric
\begin{equation}
M_n=\frac{1}{n}\sum_i^nB_iP_iS_iW_i\frac{1}{10},
\end{equation}
is needed to exclude that possibility.  Since this is the average of the validation metrics, multiple poor scores give a poor average.  Therefore it can be interpreted that a score below 0.3 is considered poor, a score in the range 0.3 to 0.7 is considered reasonable, and a score above 0.7 is considered good.  Like the first metric, this metric is also susceptible to the possibility of yielding an artificially good score.  For the second metric, selecting only a single measure of outstanding quality could produce a good score.  However this selection would not produce a good score in the first metric.  Consequently both metrics must be considered jointly.  The two metrics can be combined as components of a final metric vector ${\bf V}=(M_s, M_n)$, with a high score on both needed for confidence in the validation.  
This sample composite metric does not include any inputs directly assessing qualification, but could easily be modified to do so, following the pattern used for primacy hierarchy and sensitivity.  For example, a new factor could be inserted that lowers the metric as model parameters move away from a region of validity.
Because composite metrics like the one described above have not been tested in actual comparisons of models and experiments, we expect that composite metrics will have to be developed, modified and refined through a process of trial and error.

\medskip

\noindent{\bf IV. BUILDING A CULTURE OF VALIDATION}

\medskip

In the past, questions of code fidelity have been narrowly viewed as the exclusive province and responsibility of modelers.  The temptation to apply this attitude to validation wholly misunderstands what validation means and represents.  Validation can only be carried out through a close interaction between experiment, theory and modeling, with all sides actively participating in a coordinated fashion.  Moreover, modeling will play such an important role in designing and executing fusion discharges that a validated predictive model could be said to be the end product of the expertise and understanding of the US fusion sciences effort over the next decade or so.  Achieving this product will require building a culture of validation.  

The close coordination between experiment, theory and modeling will mean that validation campaigns are conceived, designed and executed as joint exercises between experimental observation and model solution.   It is understood that codes subjected to this validation activity will have been first rigorously verified.  The joint nature of the exercise will ensure that units, conventions, and definitions used in modeling and experiment are equivalent.  The experiments will be conceived and designed as a hierarchy progressing from simplest physics and geometry to more complicated cases.  Experiments executed for validation will not likely showcase fusion performance for the device.  Nonetheless they must be incorporated in long term planning. Their importance must be recognized and promoted from the highest levels of management, and runtime must be allocated accordingly.  The design of experiments must account for the outcomes of code qualification, the assessment of uncertainties and deficiencies, the primacy hierarchy, the sensitivity landscape, and the way the validation metric will be applied.  These procedures not only evaluate the realities of accuracy, resolution, regimes, parameter ranges, quantities measured and in what way, but they will dictate choices made when options are available.  It will be important to operate the codes in a predictive mode, well ahead of the experiment.  Blind and double blind comparisons should become the norm.  Full disclosure of difficulties, and shortcomings in comparisons should be made and reported.  The activity of validation and the creation of meaningful and credible validation metrics are more important than a perceived favorable outcome of a comparison.  Healthy scientific skepticism about favorable results should not be held in abeyance.  
Indeed, composite validation metrics based on worst comparisons might be developed as a way of checking if a single, well supported discrepancy might be enough to invalidate the whole basis for a model, even when other quantities are seemingly predicted with fidelity.
It is important that this type of activity be recognized and rewarded as valid and worthwhile for individuals building a career.  Journals are taking note of the crucial nature of verification and validation.  This is exemplified in the recent editorial statement of the {\it Physics of Plasmas} \cite{POP}, which states that ``it is the policy of {\it Physics of Plasma}s to encourage the submission of manuscripts whose primary focus is the verification and validation of codes and analytical models aimed at predicting plasma behavior''.  Verification and validation should also be appropriately represented in meetings and not overlooked in invited talk selection.

The design of advantageous validation experiments is not just a planning question for existing facilities, but is a community-wide planning question for future facilities and facilities development.  Special experimental conditions can remove complicating factors, uncertainty, and deficiency.  This will typically require the design and execution of dedicated experiments that systematically scan relevant parameters and provide the necessary conditions to facilitate diagnostic measurements (e..g, profile and fluctuation data). The value of a validation experiment will be significantly enhanced if simulations and models can reproduce not only single-point comparisons across the primarcy hierarchy, but also the variation and trends with the relevant variables (e.g., normalized gyroradius, collisionality, beta, safety factor). Performing model calculations a priori to help guide the design of experiments is especially useful for determining which parametric variations yield the greatest value in terms of distinguishing and validating models. Such experiments can probe lower primacy levels and favorably treat sensitivity.  Such conditions can be sought in experiment design and utilization.  For validation purposes experiments should be sought that offer simplified geometry or magnetic topology, that freeze key quantities that otherwise vary, that set parameters in regimes of simpler physics, that integrate fewer disparate effects, or that provide enhanced diagnostic access and capability.  The same kind of strategically opportunistic thinking can be applied to diagnostics and analysis techniques.  Diagnostic developments that increase sensitivity, improve resolution, access new spectral regimes, allow measurement of more fluctuating fields, retain and utilize phase data, {\it etc.}, will enhance validation.  Analysis techniques that can expand experimental access to different levels in the primacy hierarchy or modify sensitivity, such as bispectral deconvolution analysis \cite{kim} and fractional derivatives \cite{fracder}, are also valuable.

\medskip

\noindent{\bf V. CONCLUSION: BEST PRACTICES IN MODEL ASSESSMENT}

\medskip

Validation is designed to confront the challenges and potential limitations intrinsic to numerical modeling of highly complex, nonlinear physical systems.  In magnetically confined fusion plasmas there are severe constraints and difficulties associated with limitations in measurement capability, strong sensitivities in physical behavior, and extreme demands on modeling that have lead to an array of physically distinct, non overlapping models.  To handle these challenges validation approaches have been described herein.  Some of these form part of the standard canon of validation.  Others have been introduced, emphasized, or modified here, including the primacy hierarchy, sensitivity analysis, and the composite validation metric.   In the practice of fusion modeling, validation is presently an ideal, not a reality.  Indeed, the necessary coordination between experiment, modeling and theory; the desirability of designing and conceiving experiments expressly for validation; and the need for objective assessment, make validation an activity of the whole of the fusion research enterprise, not merely the part devoted to modeling.  It is not likely that significant new funding will permit a separate validation industry running in parallel with existing experimental and modeling efforts.  However, because the types of challenges in fusion validation are strongly rooted in physics, validation can become an intrinsic part of how magnetic fusion research is carried out, as suggested in the prior section.

The shift to a culture of validation will not occur instantaneously.  Stimulus for making such changes will gain momentum as validation efforts become more widespread.  To that end we recommend that the following four steps be implemented and publicly documented. 

\smallskip

\noindent{Step 1}:  Qualification - Describe the assumptions going into a model and its region of applicability.  The more explicitly the region is defined (even including conditions outside the region) the better.

\smallskip
 
\noindent{Step 2}:  Verification - Describe steps taken to ensure proper solution of the numerical model.  At the minimum this should include convergence tests in time, space and particle number.
 
 \smallskip
 
\noindent{Step 3}:  Validation Part I - Construct and describe a primacy hierarchy and validation metrics for the model and its validation with experiment.  Do sensitivity analysis.  Calculate the validation metrics, explaining, if possible, the physics behind disagreements and agreements. 
 
 \smallskip
 
\noindent{Step 4}:  Validation Part II - Construct a composite metric and quantify its ``goodness''.  Show that the problem range is in the qualified region (or, if not, explain why this is reasonable).

\smallskip

\noindent{Following} these steps will foster confidence in validation efforts. It will ensure proper use of models, and through public documentation, will reduce the possibility that a model is used outside its range of applicability.  
As these activities become more widespread a better picture will emerge of matters like the sensitivity landscape and the primacy hierarchy, and how they affect validation.  This, in turn, will drive refinements in the use and implementation of validation metrics, and their interpretation in prediction.

\bigskip

\noindent{\bf ACKNOWLEDGMENTS}

\medskip

This work was supported by the Office of Fusion Energy Sciences of the US Department of Energy under a directive from the US Burning Plasma Organization.

\newpage

\appendix

\noindent{\bf \large{APPENDIX A: GLOSSARY OF TERMS FOR}} 

\noindent{\bf \large{VERIFICATION AND VALIDATION}}

\bigskip

This list of terms and the associated definitions were based in part on a similar list in a draft report entitled ``Guidelines for the Validation \& Verification procedures'' [P. Strand et al., European Fusion Development Agreement, Integrated Tokamak Modelling Task Force Report EU-ITM-TF (04)-08], which in turn was adapted from the AIAA ``Guide for the Verification and Validation of Computation Fluid Dynamics Simulations'' (American Institute of Aeronautics and Astronautics Report AIAA G-077-1988).

\medskip

\noindent{\bf Model}: A representation of a physical system or process intended to enhance our ability to understand, predict, or control its behavior. 

\medskip

\noindent{\bf Conceptual model}: The set of observations, mathematical modeling data, and mathematical (e.g., partial differential) equations that describe the physical system. It will also include initial and boundary conditions. 

\medskip

\noindent{\bf Qualification}: A theoretical specification of the expected domain of applicability of a conceptual model and / or of approximations made in its derivation. 

\medskip

\noindent{\bf Code}: A computer program that implements a conceptual model. It includes the algorithms and iterative strategies. Parameters for the code include the number of grid points, algorithm inputs, and similar parameters, etc. 

\medskip

\noindent{\bf Uncertainty}: A potential deficiency in any phase or activity of the modeling process that is due to the lack of knowledge.  Uncertainties can arise in the models themselves and / or in the experimental data used for validation.

\medskip

\noindent{\bf Error}: A recognizable deficiency in any phase or activity of modeling and simulation that is not due to lack of knowledge.   For example, errors may be introduced by insufficient spatial or temporal discretization in a simulation or by programming bugs in the code.

\medskip

\noindent{\bf Verification}:   The process by which the fidelity of a numerical algorithm with respect to its mathematical model is established and the errors in its solution are quantified; an exercise in mathematics and computer science.

\medskip 

\noindent{\bf Validation}: The process of determining the degree to which a model is an accurate representation of the real world from the perspective of the intended uses of the model; an exercise in physics. 

\medskip

\noindent{\bf Benchmark}: A comparison of two codes; does not, by itself, verify or validate the codes.

\medskip

\noindent{\bf Calibration}: The process of adjusting numerical or physical modeling parameters in the computational model for the purpose of improving agreement with experimental data. 

\medskip

\noindent{\bf Measure}: Any measurable quantity in experiment for which the comparable quantity in a model is computed and compared.

\medskip

\noindent{\bf Modeling}: The process of construction or modification of a model
simulation: The exercise or use of a model.

\medskip

\noindent{\bf Prediction}: Use of a code, outside of its previously validated domain, to foretell the state of a physical system.

\medskip

\noindent{\bf Primacy hierarchy}: Ranking of a measurable quantity in terms of the extent to which other effects integrate to set the value of the quantity.  Assesses ability of measurement to discriminate between different non-validated models.

\medskip

\noindent{\bf Regression testing}: Repeating a set of previously run simulation test cases to ensure that intervening code modifications have not introduced errors into the code.
\medskip

\noindent{\bf Sensitivity analysis}: The study of how the variation in the output of a model (numerical or otherwise) can be apportioned, qualitatively or quantitatively, to different sources of variation.

\medskip

\noindent{\bf Validation metric}: A formula for objectively quantifying a comparison between a simulation result and experimental data.  The metric may take into account errors and uncertainties in both sets of data as well as other factors such as the primacy of the quantities being compared.  It may be designed either to test a hypothesis (``which of two models better matches the data?'') or to determine the accuracy of the model for the application at hand (``the differences between the code and experiment lie within the 90\% confidence interval'').

\bigskip\

\noindent{\bf \large{APPENDIX B: SIMPLE VALIDATION METRIC DETAILS}}

\bigskip

As described in the main text, the simulation-experiment comparison
shown in Fig.~4 consists of experimental data over a range
of $x$ with two or more experiments performed at each $x$.  The mean
experimental result is plotted along with the $90 \%$
confidence interval computed from these results using the expressions
in \cite{OandB}.

Simulations
have been carried out over this same range of $x$.
The availability of multiple experiments at a given $x$ allows a
more precise characterization of experimental inputs to the simulation.
If a single simulation is to be done, the mean of these inputs can
be used.  If a separate simulation can be carried out for each experimental
point, the mean of the simulation results can be used in the
comparison.
In the latter case, a distribution of
simulation results about the mean is also obtained.  
In some cases, 
one might wish to assign a width to the distribution of the simulation
results associated with
unknown parameters or due to statistical uncertainties.
To simplify further discussion, we will assume that in all of these cases,
we can determine a mean simulation value $\Sigma_{m}$ and a 
variance $\sigma_{m}^2$.  These are plotted in Fig.~4.
Determining the magnitude of this variance $\sigma_{m}^{2}$ does not
necessarily require many (potentially) expensive simulation runs.
Effective techniques for estimating this uncertainty, e.g., ``response
surface methodologies'' \cite{chen04}, have been developed for precisely
this purpose.

The interpretation of the confidence interval is
that the {\em true} experimental mean is in the interval \cite{OandB}
\[
\bar{\Sigma}_{e} - \Delta \Sigma_{CI} \rightarrow
\bar{\Sigma}_{e} + \Delta \Sigma_{CI}.
\]
In this way, the true error in the simulation result, in the limit
of $\sigma_{m} \ll \Sigma_{m}$, is inferred to be in the interval
\[
(\Sigma_{m} - \bar{\Sigma}_{e}) - \Delta \Sigma_{CI} \rightarrow
(\Sigma_{m} - \bar{\Sigma}_{e}) + \Delta \Sigma_{CI}.
\]

To interpret the resulting data shown in Fig.~4 in a quantitative
way, we examine the average of the absolute value of the 
difference between the experimental mean 
and the simulation results over the
distribution of the latter.  One obtains an expectation value of
the mean error as approximately
\[
{\cal E}(\Delta \Sigma_{e}) \simeq \max \left[ |\Sigma_{m} - \bar{\Sigma}_{e}|,
\sqrt{2/\pi} \sigma_{m} \right].
\]
This is plotted in Fig.~5 
along with the (positive) confidence interval.
This is then the local validation metric.

The effectiveness of the model, and of the validation exercise itself, can
be directly assessed when the data are presented in this manner.
For example, the simulation error exceeds the $90 \%$ confidence
interval at the largest $x$ values, where the confidence intervals
are relatively small, suggesting that future model improvements would
be most profitably focused on this regime.  In contrast, the confidence
intervals are relatively large for small to medium values of $x$,
preventing the trend predicted by the simulation in Fig.~4
from being tested.  The upshot of this exercise would then be that more experiments
and / or more precise measurements are needed in this regime.

A global metric can be computed by normalizing the mean error by the
experimental mean and then averaging over $x$ \cite{OandB},
\[
\left | \frac{\Delta \Sigma_{e} }{\bar{\Sigma}_{e}} \right | = 
\frac{1}{x_{\rm max} - x_{\rm min}} \int_{x_{\rm min}}^{x_{\rm max}}
\frac{ \max \left [ |\Sigma_{m} - \bar{\Sigma}_{e}|,
\sqrt{2/\pi} \sigma_{m} \right]}{ | \bar{\Sigma}_{e} | }.
\]
To interpret this, we need a corresponding averaged confidence
interval \cite{OandB},
\[
\left | \frac{\Delta \Sigma_{CI}}{\bar{\Sigma}_{e}} \right | = 
\frac{1}{x_{\rm max} - x_{\rm min}} \int_{x_{\rm min}}^{x_{\rm max}}
\frac{ | \Delta \Sigma_{CI} |}{ | \bar{\Sigma}_{e} | }.
\]
The result of this computation is a statement along the lines of ``the
average relative error is $23 \% \pm 15 \%$ with $90 \%$ confidence''.
Initial validation attempts may frequently yield large values for
both the average relative error and confidence intervals.  In this
case, the global metric is of less benefit than the local metric. 
Namely, one would use the latter as an indication of where to best focus
future efforts.

In the case where both the average relative error and confidence intervals
are small, one then needs to consider whether this level of 
accuracy is adequate for the intended applications of the
model.

\bigskip

\noindent{TABLE CAPTION}

\bigskip

\noindent{Table 1:} Scorecard of validation approaches for resolving common challenges.

\bigskip

\noindent{FIGURE CAPTIONS}

\bigskip

\noindent{FIG. 1:} Primacy hierarchy for particle transport.

\bigskip

\noindent{FIG. 2:} Primacy hierarchy for power density spectrum.

\bigskip

\noindent{FIG. 3:} Comparison of model results with experiment at two levels in the primacy hierarchy.  The density fluctuation has a lower primacy level than the flux, but its discrepancy with experiment is larger.  The flux also illustrates the strong sensitivity of transport to the gradient scale length, showing a factor of 5 increase in the flux for a 60\% increase in gradient scale length.

\bigskip

\noindent{FIG. 4:} Generic depiction of a comparison of experimental simulation data with $\pm$90\% confidence intervals computed for the former and an uncertainty of $\sigma_m$ shown for the latter.

\bigskip

\noindent{FIG 5:}  Comparison of the expectation value of the average error between the experimental and simulated data (in blue) with the 90\% confidence interval (in red).

\bigskip

\noindent{FIG 6:}  Taylor diagram for Fig. 3 of Rhodes et al. \cite{rhodes}, which displays the closeness of ion temperature gradient models to experiment (labeled Ref).  In this polar plot the radial variable is standard deviation normalized to the mean and the angle (measured from the vertical axis) is the inverse cosine of the correlation coefficient.

\newpage

\begin{table}
\resizebox{\textwidth}{!}{
\includegraphics{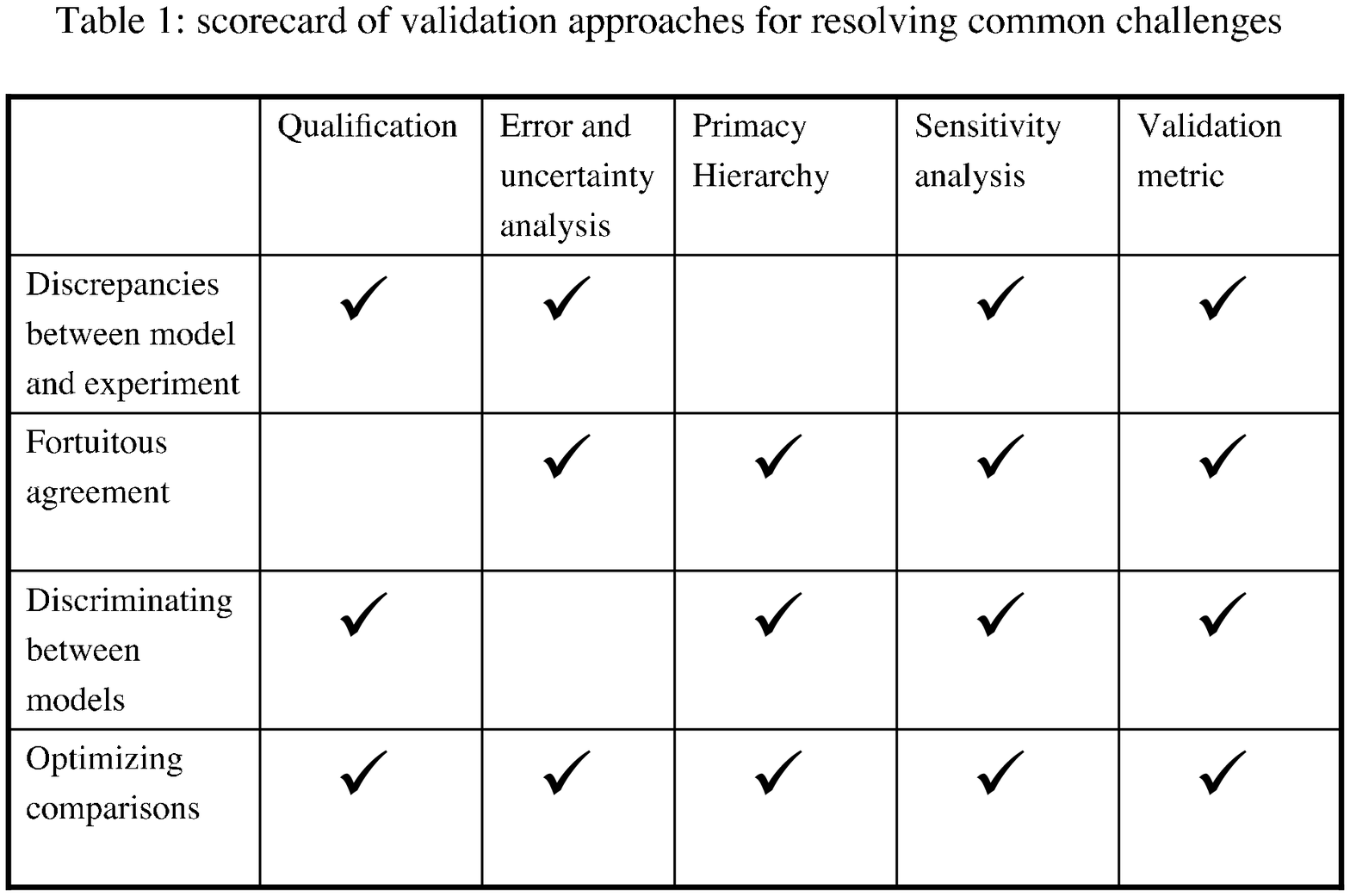}}
\caption{}
\label{fig1}
\end{table}

\newpage

\begin{figure}
\resizebox{\textwidth}{!}{
\includegraphics{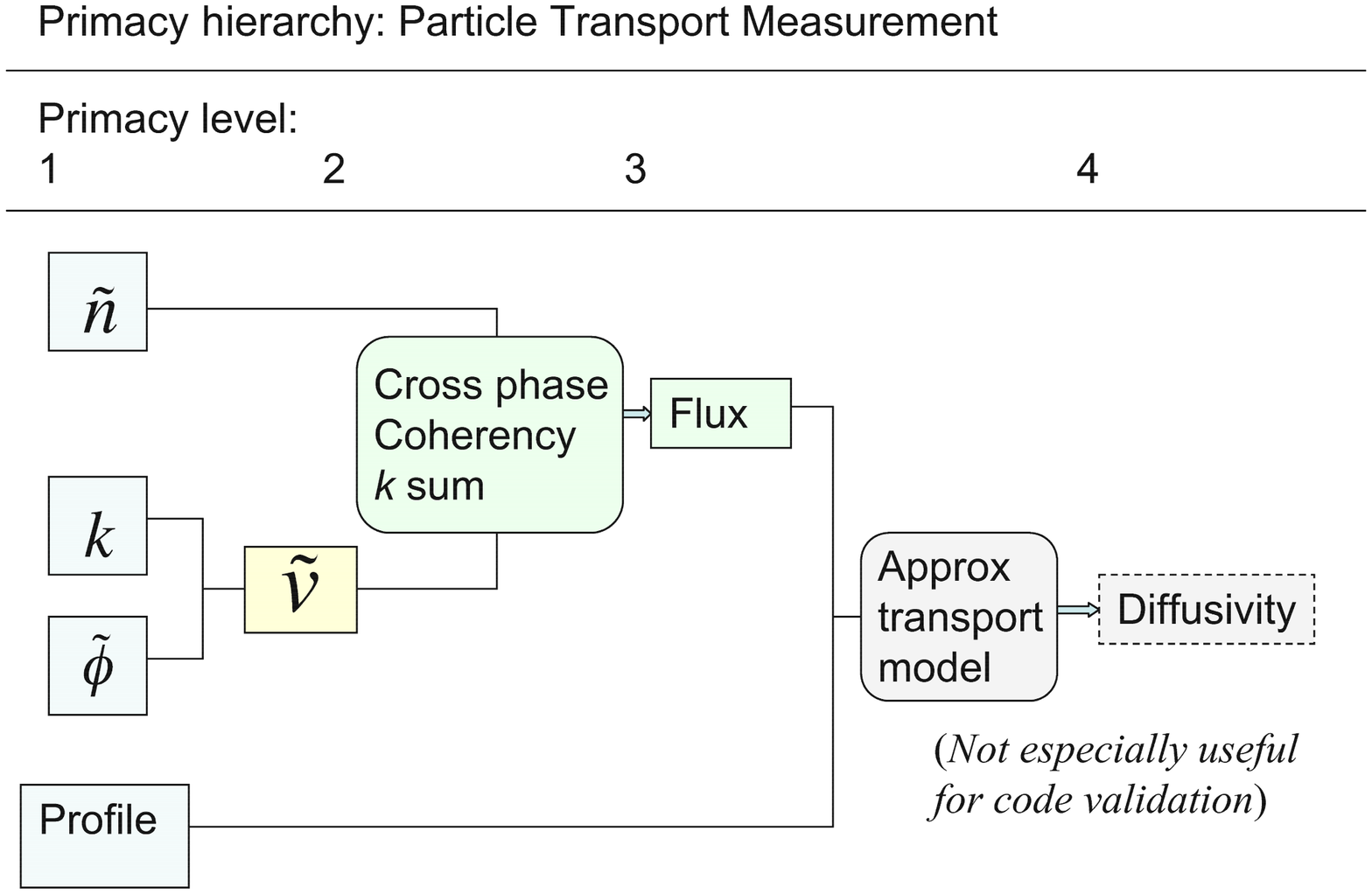}}
\caption{}
\label{fig1}
\end{figure}

\newpage

\begin{figure}
\resizebox{\textwidth}{!}{
\includegraphics{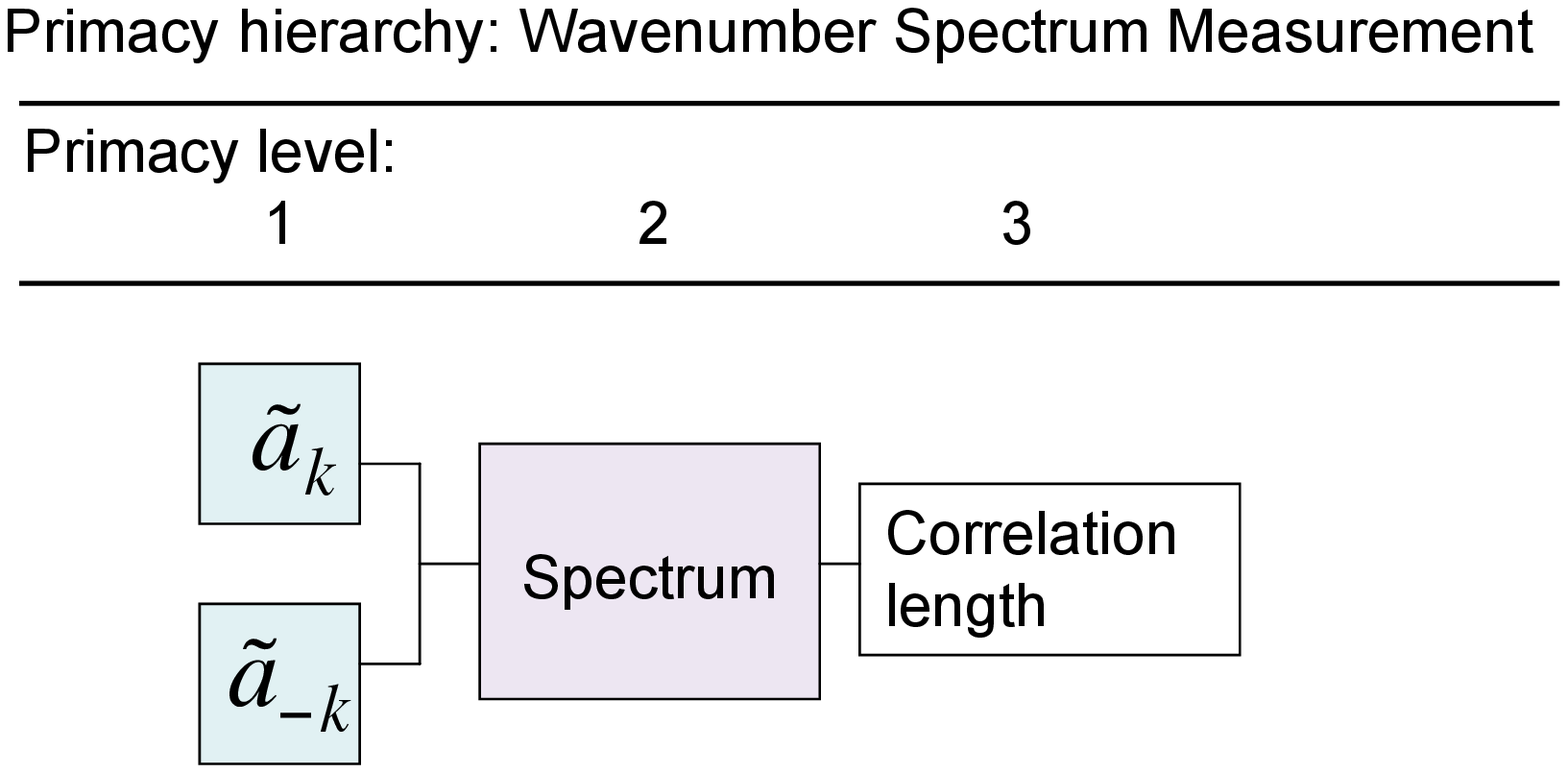}}
\caption{}
\label{fig1}
\end{figure}

\clearpage
\newpage

\begin{figure}
\includegraphics[width=18.0cm]{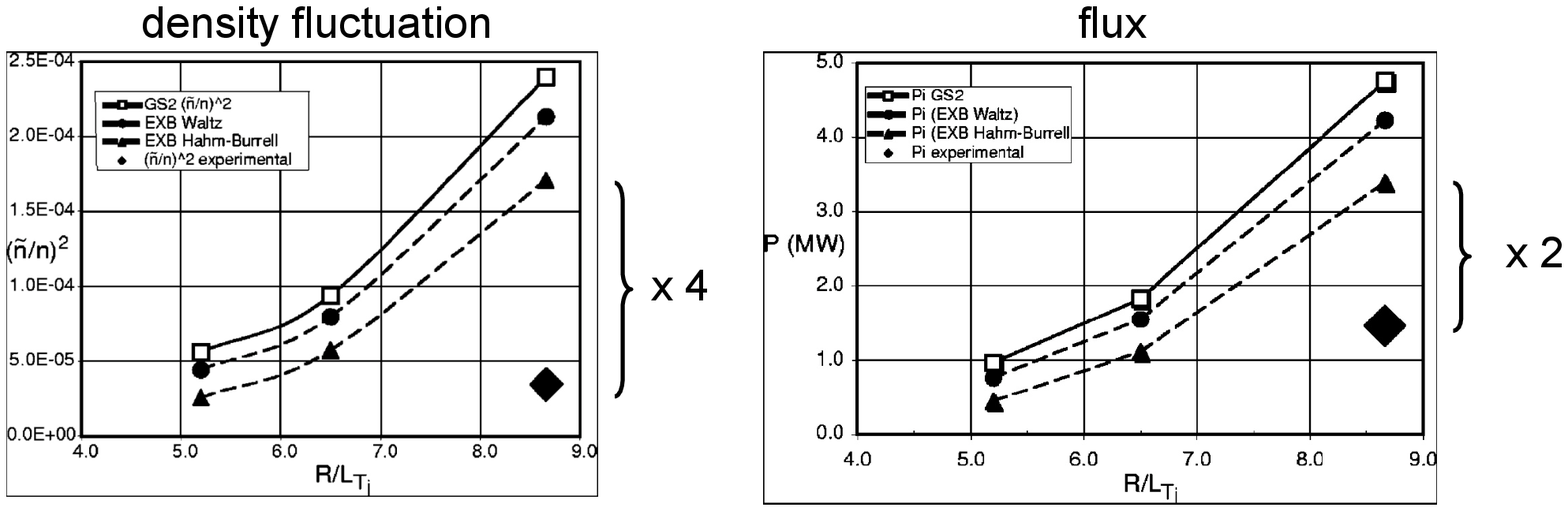}
\caption{}
\label{fig1}
\end{figure}

\clearpage
\newpage

\begin{figure}
\resizebox{\textwidth}{!}{
\includegraphics{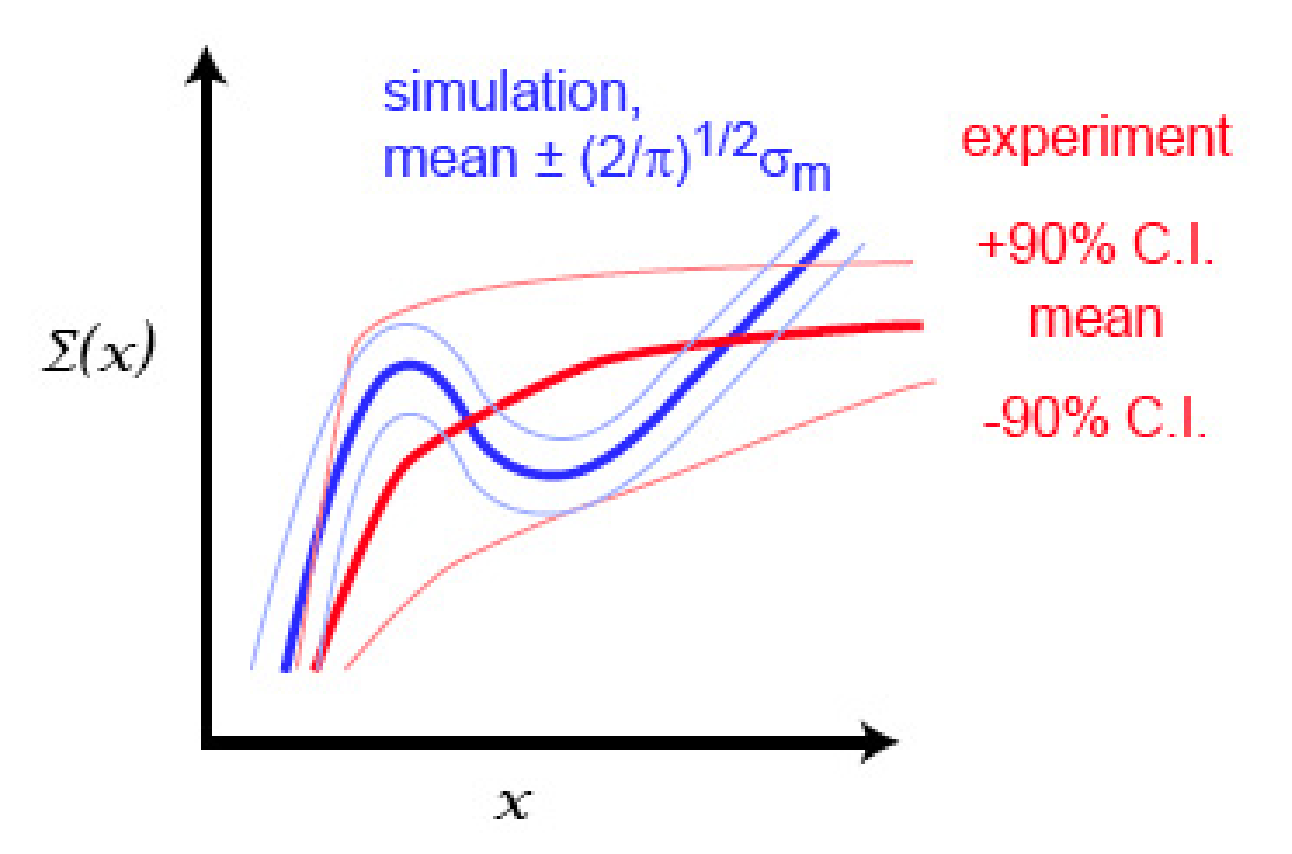}}
\caption{}
\label{fig2}
\end{figure}

\newpage

\begin{figure}
\resizebox{\textwidth}{!}{
\includegraphics{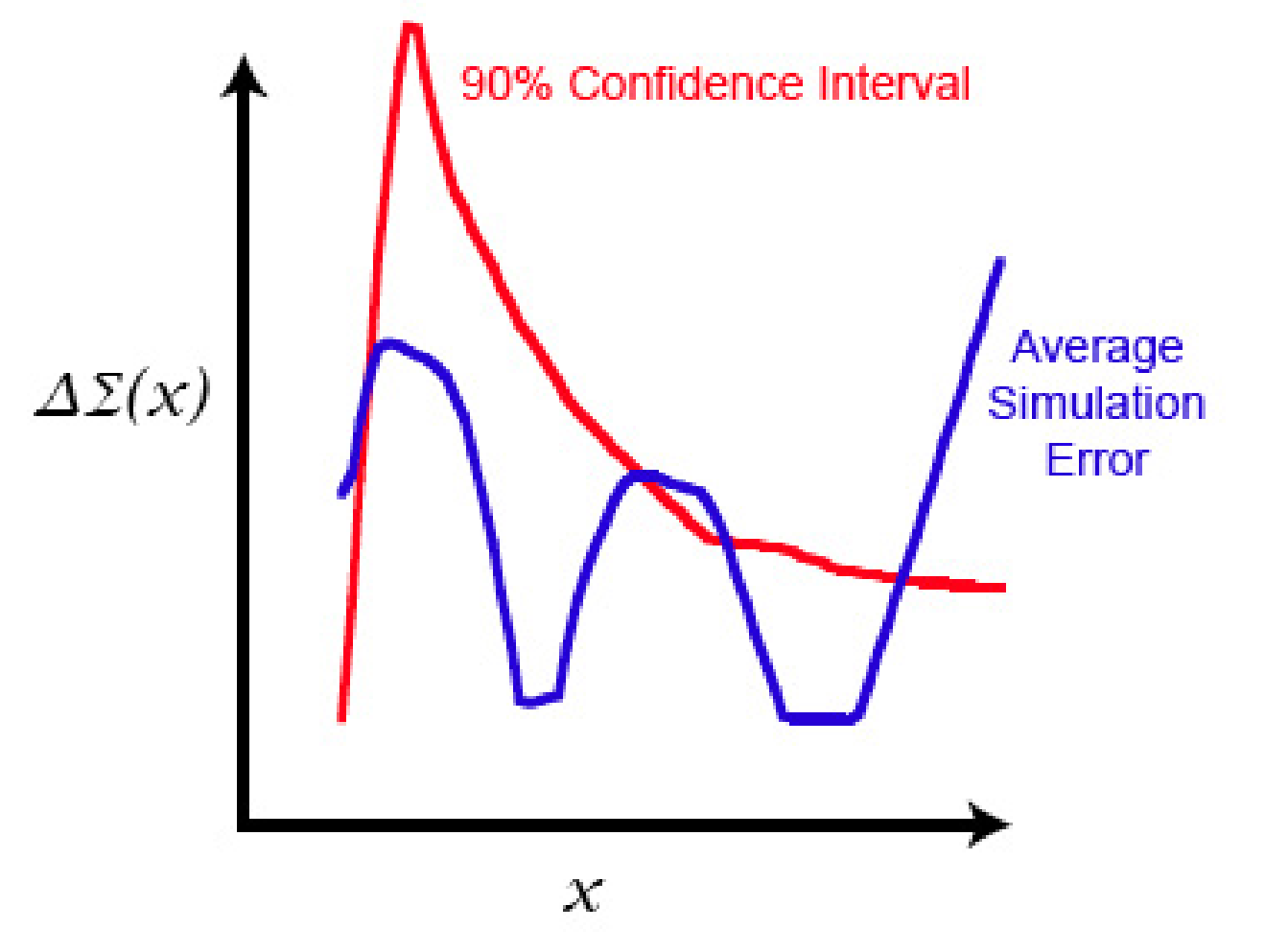}}
\caption{}
\label{fig2}
\end{figure}

\newpage

\begin{figure}
\resizebox{\textwidth}{!}{
\includegraphics{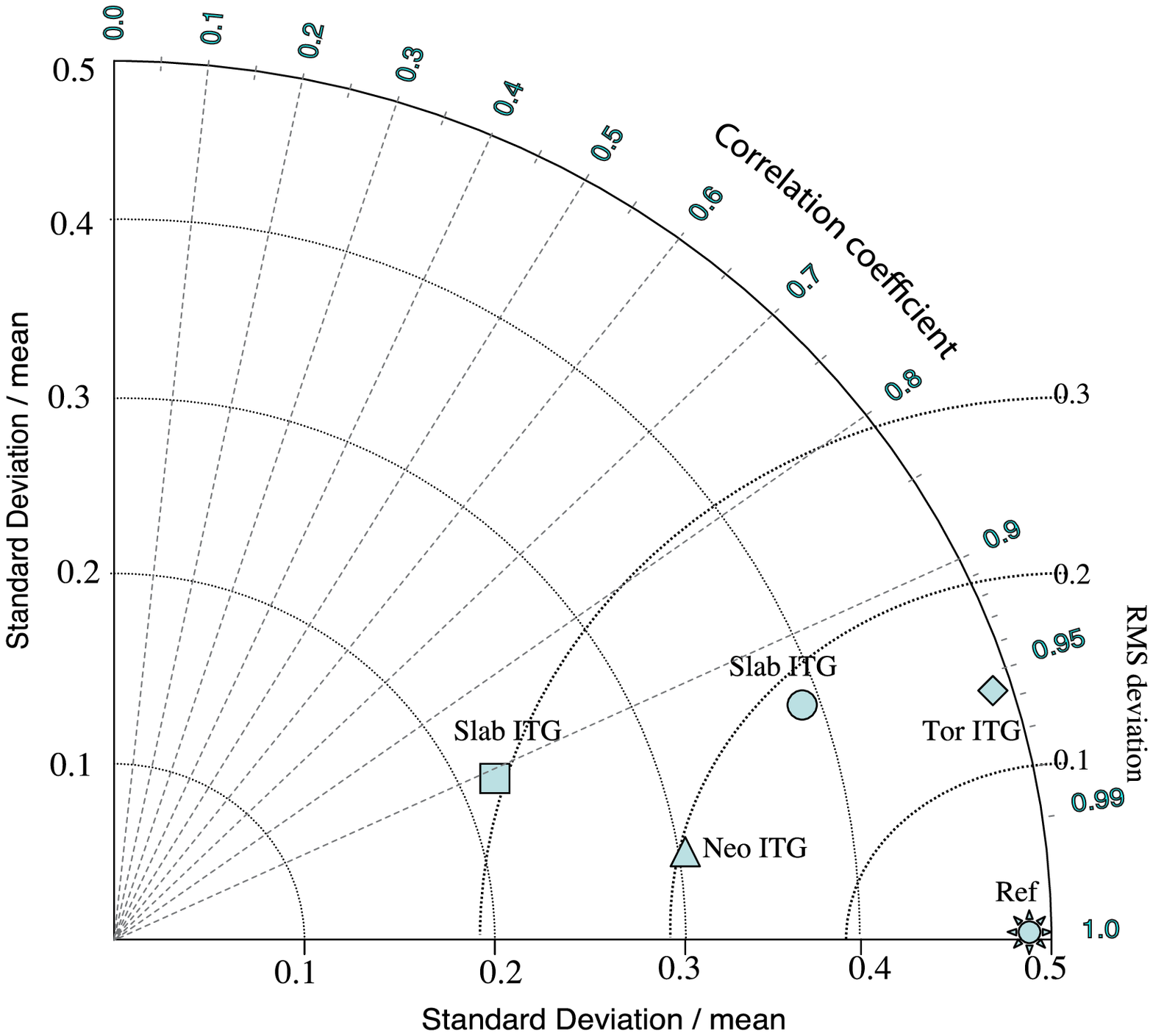}}
\caption{}
\label{fig2}
\end{figure}

\end{document}